\documentclass[
 reprint,
 amsmath,amssymb,
 aps,
 prl
]{revtex4-2}

\usepackage{graphicx}% Include figure files
\usepackage{dcolumn}% Align table columns on decimal point
\usepackage{bm}% bold math
\usepackage{gensymb}
\usepackage{xcolor}
\usepackage{array,multirow,lipsum}
\usepackage{wasysym}

\renewcommand{\figurename}{\textbf{Fig.}}
\renewcommand{\thefigure}{\textbf{\arabic{figure}}}

\begin{document}

%\preprint{APS/123-QED}

\title{A material-agnostic platform to probe spin-phonon interactions using high-overtone bulk acoustic wave resonators}% Force line breaks with \\

\author{Q. Greffe}
 \email{quentin.greffe@neel.cnrs.fr}
\author{A. Hugot}
\author{S. Zhang}
\author{J. Jarreau}
\author{L. Del-Rey}
\author{E. Bonet}
\author{F. Balestro}
\author{T. Chanelière}
\author{J. J. Viennot}
 \email{jeremie.viennot@neel.cnrs.fr}
\affiliation{Univ. Grenoble Alpes, CNRS, Grenoble INP, Institut Neel, 38000 Grenoble, France
}%

\date{\today}% It is always \today, today,

\begin{abstract}
\textbf{Spin-phonon interactions have a dual role in emerging spin-based quantum technologies. While they can be a limitation to device performance through decoherence, they also serve as a critical resource for coherent spin control, detection, and the realization of spin-based quantum networks. However, their direct characterization remains a challenge and is usually material-dependent. Here, we introduce a technique to probe spin-phonon coupling at millikelvin temperatures and gigahertz frequencies, using high-overtone bulk acoustic wave resonators (HBARs) integrated with arbitrary crystals via visco-elastic transfer of thin-film lithium niobate transducers. By tuning the Larmor frequency of dilute spin ensembles into resonance with HBAR modes, we extract the anisotropy and strength of spin-phonon interactions from acoustic dispersion and dissipation measurements. We demonstrate this approach in calcium tungstate (CaWO$_4$) and yttrium orthosilicate (Y$_2$SiO$_5$), achieving cooperativities up to 0.5 for erbium dopant ensembles. Our method enables the study of spin-phonon interactions in complex crystalline materials, with minimal fabrication constraints. These results will facilitate the design of hybrid quantum systems and the quest for ion-matrix combination with enhanced spin-phonon coupling.}
\end{abstract}   

\maketitle

Spin-phonon interactions play a dual role in spin-based quantum technologies: while they can limit device performance through decoherence, they also serve as a critical resource for coherent spin control and quantum transduction. However, their direct and material-agnostic characterization has remained elusive.

Spin–phonon coupling has long been investigated primarily through the lens of decoherence, omnipresent at room temperature. Its study has both driven and been propelled by advancements in low-temperature physics. While cryogenic operation mitigates phonon-induced decoherence, phonons remain central to spin dynamics, as spin–lattice relaxation often governs the population decay. The investigation of spin-phonon interactions originated in Electron Paramagnetic Resonance (EPR), in order to understand the spin relaxation mechanisms \cite{Abragam1970, Larson1966}. The longitudinal relaxation time $T_1$, when limited by spin-lattice relaxation, provides a measure of the average coupling between spins and the phonon bath. However, such measurements offer only indirect access to spin–phonon interactions. Direct experimental probes of spin–phonon coupling remain scarce, and even fewer studies have reported the reciprocal effect of spins on acoustic properties. Early efforts addressed this challenge by using acoustic waves, rather than microwaves, to excite and detect spin ensembles in acoustic paramagnetic resonance (APR) experiments \cite{Tucker1961,Altshuler1962,Joffrin1965,Dobrov1966}. These pioneering experiments employed quartz rods within 3D microwave cavities to generate and detect GHz-frequency acoustic waves \cite{Jacobsen1959}. Yet, this approach faced significant technical challenges, including the difficulty of achieving reproducible mechanical contacts between large quartz crystals and the materials under investigation. Additionally, the bulky nature of these setups rendered them incompatible with integration into superconducting 3D vector magnets and dilution refrigerators.

The advent of quantum technologies, spanning quantum sensing \cite{Degen2017, Aslam2023, Budakian2024} and quantum information processing \cite{Bradley2019, Wolfowicz2021, OSullivan2025}, has catalyzed a paradigm shift. Phonons are no longer considered as a perturbation of spins, but rather as a valuable resource for coherent control and readout of qubits. This transformation has been driven by advances in nanomechanical and acoustic systems operating in the quantum regime \cite{Barzanjeh2021,Chu2020}. Acoustic waves are now used for the coherent manipulation of spins \cite{Lee2017,Whiteley2019,Maity2022,Dietz2023}, and beyond that, they have the tantalizing potential to realize analogs of circuit quantum electrodynamics using single spins and mechanical modes \cite{Lee2017,Lemonde2018}. A recent milestone in this direction is the observation of the acoustic Purcell effect with a single color center in diamond \cite{Joe2025}. 

In this context, the study of spin-phonon interactions has regained central importance. The wide diversity of spin defects and crystalline environments poses a significant challenge for the consistent and general characterization of spin-phonon coupling, both theoretically and experimentally. Moreover, spin defects with strong spin-phonon coupling often exhibit significantly reduced lifetimes, even at cryogenic temperatures as low as 4 K. These challenges underscore the need for an experimental technique compatible with diverse materials and with standard dilution refrigerator infrastructure. \\

In this Article, we introduce a material-agnostic technique to probe, at GHz frequencies and millikelvin temperatures, the coupling between dilute spin ensembles and high-overtone bulk acoustic wave resonators (HBARs). Our method employs prefabricated thin-film piezoelectric transducers made of lithium niobate, which we transfer onto mm-sized samples of arbitrary polished materials. In crystals hosting paramagnetic spins, we use an external vector magnetic field to tune the Larmor frequency of the spins in resonance with the HBAR modes. From the acoustic dissipation and dispersion, we extract the strength and the anisotropy of the spin-phonon coupling. We demonstrate this technique on two materials, CaWO$_4$ and Y$_2$SiO$_5$, doped with rare earth ions and in particular erbium. CaWO$_4$ is a well-characterized material, with high symmetry, and with existing data that we compare to our findings. Y$_2$SiO$_5$ is a well-studied but more challenging material, with lower symmetry, and a lack of experimental data on spin-phonon coupling, illustrating the advantage of our approach. For 20 ppm of Er$^{3+}$ in CaWO$_4$, we achieve cooperativities on the order of 0.5 between the spin ensemble and the HBAR modes. \\

\begin{figure*}[ht]
\includegraphics{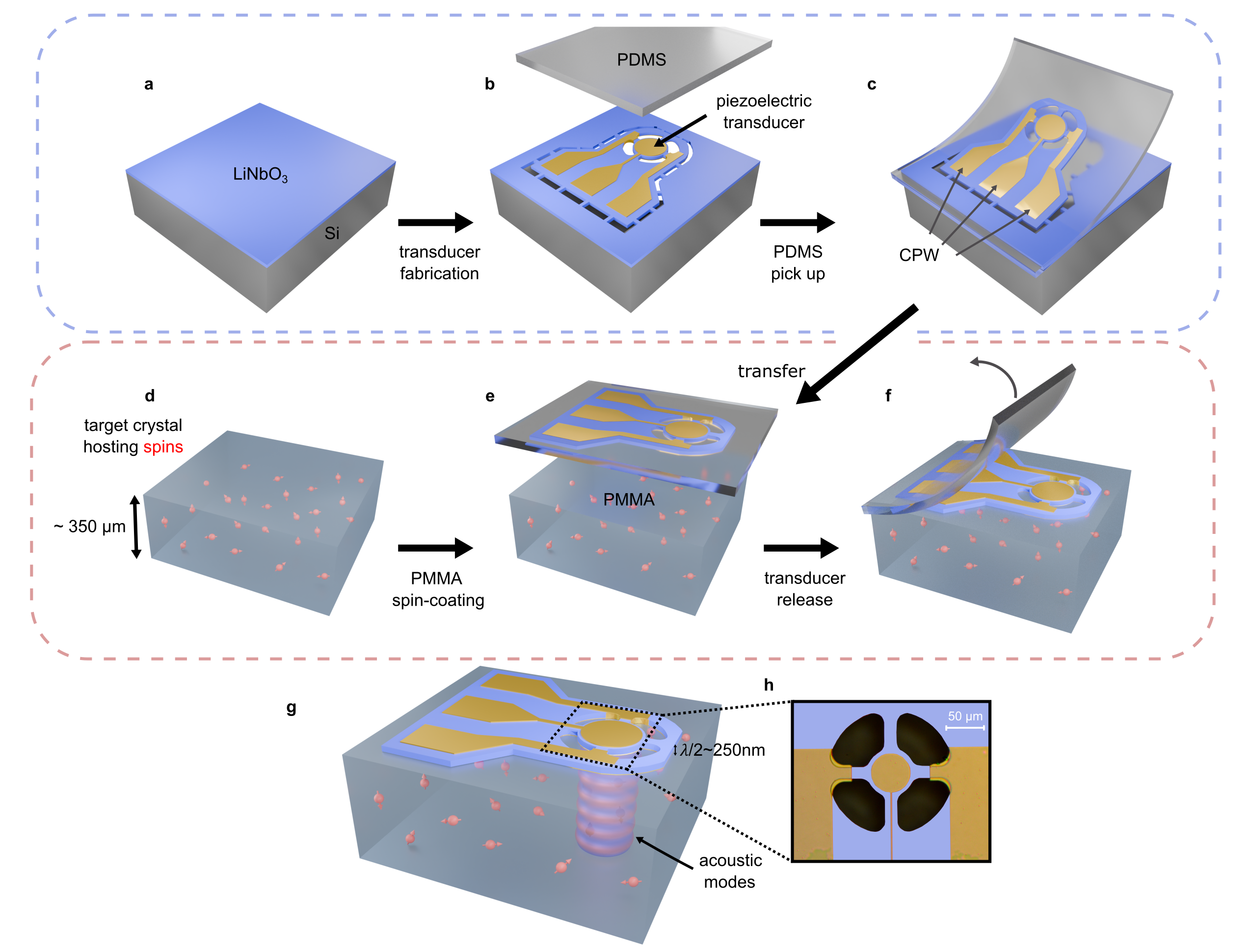}% Here is how to import EPS art
\caption{ \label{Fig1} 
\textbf{Manufacturing High-Overtone bulk Acoustic wave Resonators (HBARs) in arbitrary crystals.} 
\textbf{a,} The fabrication of the piezoelectric transducer starts from thin films of LiNbO$_3$ on Si. \textbf{b,} The piezoelectric transducer, patterned in LiNbO$_3$, is fitted with top and bottom metallic electrodes, which are respectively connected to the signal and ground pads of a coplanar waveguide (CPW). The entire structure (transducer and CPW) is free standing and attached to the carrier Si substrate by LiNbO$_3$ tethers. \textbf{c,} We use a PDMS stamp to pick up the entire structure from the Si substrate. \textbf{d,} A target crystal (hosting spins) must have two parallel faces polished. \textbf{e.} After spin-coating a thin layer of PMMA ($\approx$ 60 nm) on the crystal, we heat it up to 150 \degree C and we transfer the piezoelectric structure onto it. \textbf{f,} We retract the PDMS stamp and leave the piezoelectric structure bonded onto the crystal. \textbf{g,} We excite and detect the acoustic modes in the crystal via the bonded piezoelectric transducer. \textbf{h,} False-colored optical micrograph of a transducer bonded onto a transparent crystal, showing regions where the bottom electrode is connected to the electrical ground of the CPW. }
\end{figure*}

\begin{figure*}[ht]
\includegraphics{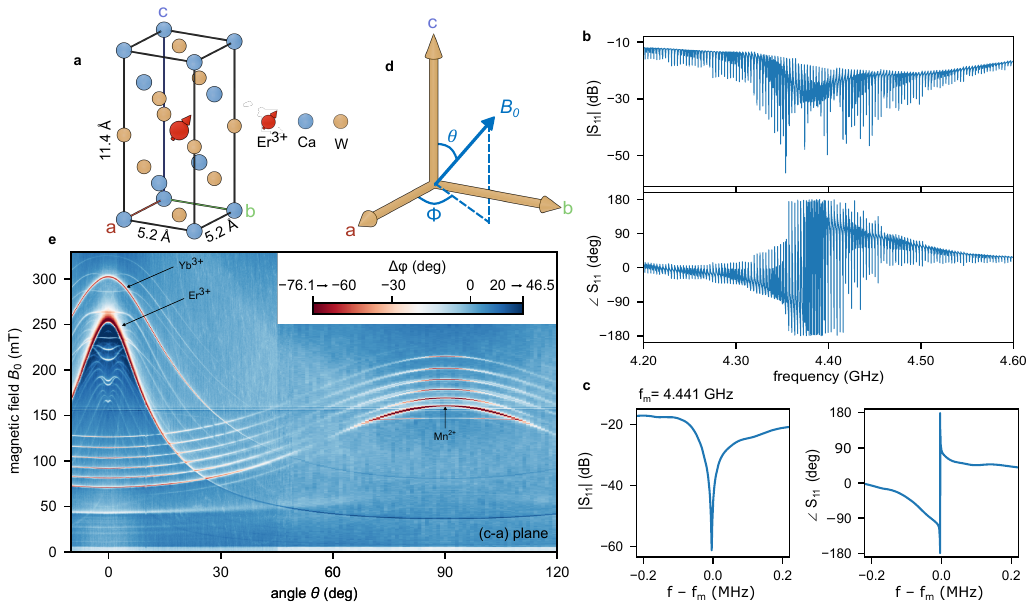}% Here is how to import EPS art
\caption{
\label{Fig:CaWO$_4$} 
\textbf{Acoustic electron paramagnetic resonance in calcium tungstate (CaWO$_4$) using an HBAR.}
\textbf{a,} Unit cell of the tetragonal lattice of CaWO$_4$, with one Er$^{3+}$ dopant. Oxygen atoms are omitted for clariry. \textbf{b,} Microwave reflection $S_{11}$ (magnitude and phase) of a CaWO$_4$ HBAR at low temperature and large microwave power ($\approx -72$~dBm at the HBAR input). \textbf{c,} Microwave reflection of a critically-coupled HBAR mode, at $f_m=4.441$ GHz. \textbf{d,} Spherical coordinates for the external magnetic field, with amplitude $B_0$ and orientation given by $\theta$ and $\varphi$ with respect to the crystal axes of CaWO$_4$. \textbf{e,} Relative phase shift of the microwave reflection measured on the HBAR resonance shown in \textbf{c} (color scale), as a function of the external magnetic field amplitude $B_0$ and angle $\theta$ (at $\phi=0\degree)$. The main paramagnetic resonance signal for erbium, yterbium and manganese ions are indicated. The secondary lines with weaker contrast correspond to hyperfine resonances, see Extended Data Fig. \ref{Fig:ZoomCaWO$_4$-c-axis} for more details.}
\end{figure*}

\noindent \textbf{HBARs in arbitrary crystals}\\
In order to probe, at dilution refrigerator temperatures, resonant interactions between polarized spin ensembles and acoustic modes without thermal excitations, we require acoustic modes at GHz frequencies. In mm-sized crystals, HBARs can be operated at such frequencies \cite{Zhang2006,Chu2017,Kervinen2018,Gokhale2020}. HBARs are essentially acoustic Fabry-Perot resonators, where flat mirrors can simply be realized by polished faces. To couple microwave signals into HBARs, it is common to use a piezoelectric transducer microfabricated on one of the polished faces of the crystal. This technique works well on standardized materials such as quartz, sapphire, silicon or silicon carbide \cite{Zhang2006,Chu2017,Kervinen2018,Gokhale2020}. But crystals hosting interesting spin species can be sensitive materials and specific microfabrication techniques would have to be developed for every new target material. To avoid this issue and manufacture HBARs in arbitrary crystals, we adopt a visco-elastic transfer strategy, illustrated in Fig. \ref{Fig1}. We start by fabricating batches of suspended LiNbO$_3$ transducers from commercial LiNbO$_3$/silicon wafers (Fig. \ref{Fig1}\textbf{a}). This fabrication involves few steps of lithography, metal deposition and etches, and is described in detail in the Methods section. At the end of this process, we obtain free-standing disks of LiNbO$_3$ (50 \textmu m diameter) with top and bottom gold electrodes, as illustrated in Fig. \ref{Fig1}\textbf{b}. The top electrode is directly connected to the center conductor of a coplanar waveguide (CPW), while the bottom electrode connects to the electrical ground of the CPW. As shown in Fig. \ref{Fig1}\textbf{c}, we use a polydimethylsiloxane (PDMS) stamp to pick up the entire structure, which typically breaks off from the handle substrate at the tethers. Before transferring this structure on the target crystal, we spin-coat the crystal with a thin layer of glue ($\approx$ 60 nm of PMMA, Fig. \ref{Fig1}\textbf{d,e}). To transfer, we first heat up the crystal to soften the glue ($\approx 150 \degree$C) and then stamp the LiNbO$_3$ structure onto it. We then retract the PDMS (Fig. \ref{Fig1}\textbf{f}) to obtain the HBAR structure shown in Fig. \ref{Fig1}\textbf{g,h}. \\

There are various criteria to evaluate the performances of HBARs, depending on the targeted application. For the purpose of detecting dilute spin ensembles, the key parameters are the intrinsic losses $\kappa_\text{int}$ of the HBAR and the external coupling $\kappa_\text{ext}$ of the HBAR to the CPW microwave line. Their sum $\kappa_\text{tot}=\kappa_\text{int}+\kappa_\text{ext}$ is equal to the total linewidth of the HBAR modes. As shown in the Methods, the spin contribution to the microwave signal reflecting off the HBAR scales approximately as $\kappa_\text{ext}/(\kappa_\text{int}^2-\kappa_\text{ext}^2)$. It is therefore crucial to control these parameters to reach a regime where $\kappa_\text{int}$ and $\kappa_\text{ext}$ have the same order or magnitude. 

The internal loss rate $\kappa_\text{int}$ of HBAR devices generally depends on various contributions \cite{Gruenke2025,Belles2026}. Some of these contributions can be estimated in simple ways for a plane-parallel Fabry-Perot (parallel flat mirrors). It includes the effects of surface roughness, parallelism of the mirrors as well as finite-size effects (Fresnel number). %(ratio between the square of transducer diameter, the crystal thickness and the acoustic wavelength).
To minimize internal losses, we carefully consider these contributions during the polishing of the crystal and the design of the transducers. However, losses such as the intrinsic acoustic attenuation of the target crystals cannot be controlled. Our strategy is therefore to aim for large external couplings instead of targeting ultra-low internal losses. \\

To reach large external coupling of the HBAR to the microwave line, we take advantage of the large piezoelectric coupling of LiNbO$_3$. In all the data shown in this Article, we use X-cut LiNbO$_3$ with a thickness of 250 nm, and we work with the lowest frequency shear mode (transverse acoustic polarization). In order to benchmark the robustness of the transfer technique shown in Fig. \ref{Fig1}, we run it on several silicon crystals with [100] orientation as an acoustic testbed. For this Si orientation, the transverse acoustic modes are degenerate and have a well known phase velocity. We also demonstrate our method on an amourphous material with measurements on glass (Borofloat). Some of the corresponding acoustic spectra are shown in Extended Data Fig. \ref{Fig:HBAR-Si}, demonstrating large $\kappa_\text{ext}/2\pi$ in the range 0.1 - 1 MHz for HBAR modes near a center frequency $f_0$. At this frequency, the acoustic wavelength is about twice the transducer thickness. While the exact values of $\kappa_\text{ext}$ around $f_0$ can change from one transfer to the next, we obtain external coupling $\kappa_\text{ext}/2\pi>$ 100 kHz around $f_0$ for more than 80 \% of the samples we measured (see Extended Data Table \ref{Table:MeasuredHBARs}). In addition, for HBAR modes with resonant frequencies arbitrarily far-detuned from $f_0$, we find correspondingly arbitrary low $\kappa_\text{ext}$. These values enable efficient spin spectroscopy (i.e. reaching $\kappa_\text{ext}\approx \kappa_\text{int}$) for internal acoustic quality factor $Q_\text{int} \approx 2 \pi f_0/ \kappa_\text{int}$ as low as few thousands and up to arbitrary large values. \\

After preparing a target crystal with a transducer, we package it into a copper sample holder and then insert it into a superconducting 3D vector magnet. We install this ensemble in a home-made dilution fridge with a base temperature $\approx$ 60 mK (see supplementary information). \\

\noindent \textbf{Acoustic paramagnetic resonance in CaWO$_4$} \nopagebreak

We begin our study with calcium tungstate (CaWO$_4$) crystals doped with rare earth spins at around 20 ppm. CaWO$_4$ has recently been pushed forward for advanced quantum technologies and the spin properties of rare earth dopants in this system have been studied in depth \cite{LeDantec2021,Wang2023,Chaneliere2024,OSullivan2025,Marsh2026}. The unit cell of CaWO$_4$ has tetragonal symmetry and is depicted in Fig \ref{Fig:CaWO$_4$}\textbf{a}, with an Er dopant substituting a calcium atom. This system behaves as an effective electron spin-1/2, with an anisotropic g-tensor that allows for unambiguous distinction \cite{Bertaina2007,LeDantecThesis}. Besides, the spin relaxation time of Er:CaWO$_4$ has been showed to be dependent on the magnetic field orientation  \cite{LeDantec2021,Wang2023}, indicating that the underlying spin-phonon coupling is also anisotropic.

A typical acoustic spectrum on CaWO$_4$ is shown in Fig \ref{Fig:CaWO$_4$}\textbf{b}. We observe two families of acoustic modes, with FSR 3.2 and 3.4 MHz, by analogy with optical birefringence \cite{Auld}. It arises from the anisotropy of CaWO$_4$ and the fact that we excite shear waves in the (a-c) plane, with a polarization that is aligned with neither the a nor c axes of CaWO$_4$. We find that the HBAR modes around 4.38 GHz are over-coupled ($\kappa_\text{ext}>\kappa_\text{int}$), and that $\kappa_\text{ext}$ decreases for modes away from 4.38 GHz (see Extended Data Fig \ref{Fig:Kext_CaWO$_4$}). In order to perform APR with a large contrast, we choose a mode close to critical coupling ($\kappa_\text{ext} \approx \kappa_\text{int}$) at 4.441 GHz, shown in Fig. \ref{Fig:CaWO$_4$}\textbf{c}. We then measure, at the resonance frequency of this mode, the phase of the microwave reflection coefficient as a function of magnetic field strength and orientation. When compared to the phase at zero magnetic field, the measured phase $\Delta \varphi$ is directly proportional to the acoustic dispersion induced by the spin ensemble. The results at a large input power of $-72$~dBm, for magnetic field orientations in the (a-c) plane of CaWO$_4$, are shown in Fig. \ref{Fig:CaWO$_4$}\textbf{e}. We find several lines corresponding to various spin dopants present in this sample. These lines correspond to magnetic fields for which the Larmor frequency of a particular spin ensemble matches the acoustic mode frequency. When this resonance condition is satisfied, spin-phonon interactions lead to a change in the effective sound velocity of the crystal as well as an additional absorption. Close to the c axis of CaWO$_4$ (region around $\theta=0$), we observe many resonances, corresponding to ensembles of Er$^{3+}$ spins with zero nuclear spins (even isotopes) and 7/2 nuclear spin ($^{167}$Er), as well as Yb$^{3+}$ spins with 0, 1/2 and 5/2 nuclear spins (see Extended Data Fig. \ref{Fig:ZoomCaWO$_4$-c-axis}). The six equidistant lines that slowly modulate from $\approx$ 100 mT to $\approx$ 180 mT have been previously observed using EPR techniques in similar samples \cite{LeDantecThesis} and correspond to manganese ions (Mn$^{2+}$). \\

\begin{figure*}[ht]
\includegraphics{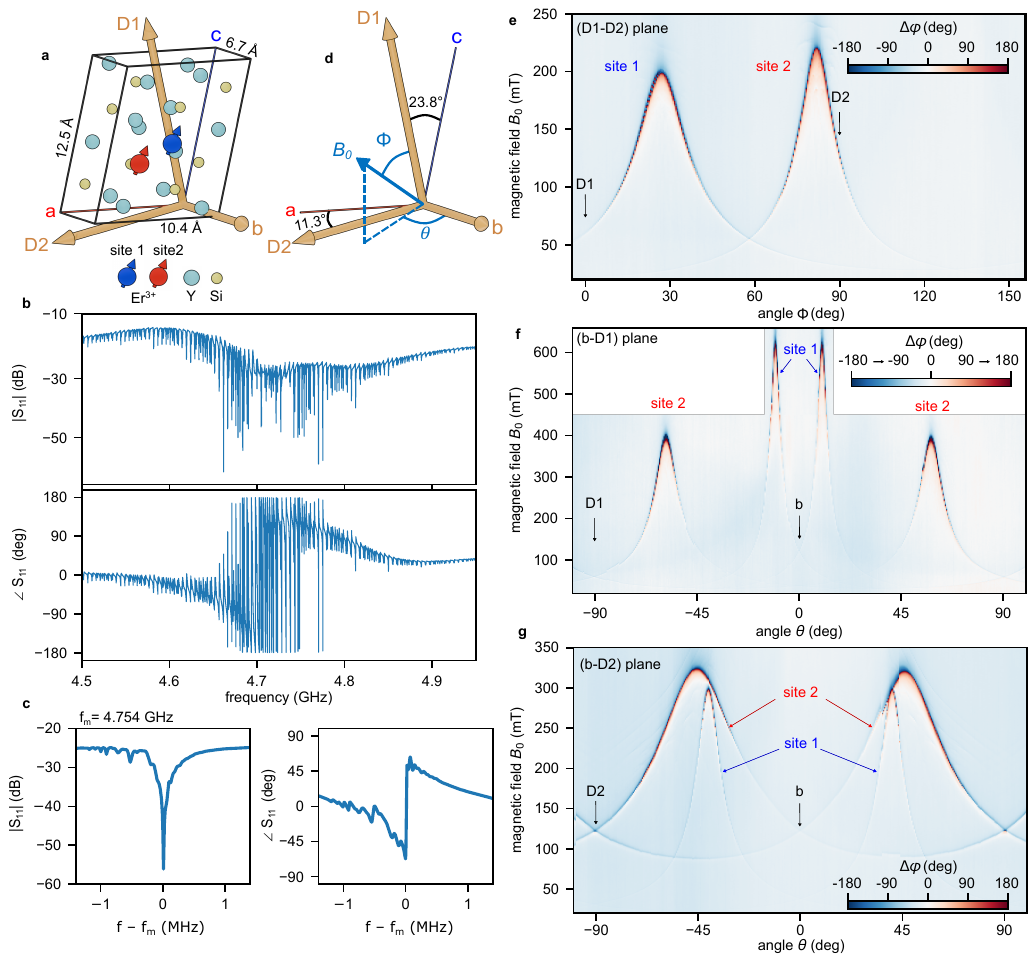}% Here is how to import EPS art
\caption{
\label{Fig:YSO} \textbf{Acoustic electron paramagnetic resonance in ytrium silicate (YSO) using an HBAR.}
\textbf{a,} Unit cell of the monoclinic lattice of YSO. Oxygen atoms are omitted for clarity. Two erbium dopants are represented, at the two independent lattice sites for substitution. The mutually perpendicular optical extinction axes (b, D1, D2), conventionally used for YSO, are represented. \textbf{b,} Microwave reflection $S_{11}$ (magnitude and phase) of an HBAR mode in YSO at low temperature. \textbf{c,} Microwave reflection of a critically-coupled HBAR mode, at $f_m=4.754$ GHz. Various parasitic modes can be seen, in particular to the left of the main acoustic resonance. \textbf{d,} Spherical coordinates for the external magnetic field, with amplitude $B_0$ and orientation given by $\theta$ and $\varphi$ with respect to the (b, D1, D2) cartesian coordinates. \textbf{e,} Relative phase shift of the microwave reflection measured on the HBAR resonance shown in \textbf{c} (color scale), as a function of the external magnetic field amplitude $B_0$ and angle $\phi$ at $\theta=90\degree$ (D1-D2 plane). \textbf{f,g} APR signal, taken as in \textbf{e} but in the b-D1 and b-D2 planes.}
\end{figure*}

\noindent \textbf{Acoustic paramagnetic resonance in YSO}\\
We move on to yttrium orthosilicate crystals (Y$_2$SiO$_5$ or YSO). YSO doped with rare earth ions is also among the leading materials for quantum applications \cite{Zhong2015,Lim2018,Dold2019,MaMa2021,Ohta2024,Chiossi2024,FWang2025}. As an yttrium compound, it has the advantage of being resilient to relatively large concentrations of rare earth dopants. The unit cell of YSO is depicted in Fig. \ref{Fig:YSO}\textbf{a}, with erbium dopants at the two different yttrium substitution sites. YSO is a monoclinic crystal and has lower symmetry than CaWO$_4$. Lower symmetry leads to highly anisotropic spin properties, and spin-phonon interactions that are more challenging to evaluate theoretically. Finding experimental ways to access these properties is therefore particularly appealing. Experimental data on spin-lattice relaxation time at low temperature is scarce \cite{Probst2013,Budoyo2018} and we have not found experimental data on the anisotropy of spin-phonon coupling.

One of the HBAR acoustic spectra we obtained on YSO doped with erbium at around 200 ppm is shown in \ref{Fig:YSO}\textbf{b}. The response is now centered around 4.7 GHz. This small frequency change can be explained by inhomogeneities in thickness of our LiNbO$_3$ thin film, which we control with a precision of around 10 \%. As shown in Fig. \ref{Fig:YSO}\textbf{b, c}, the spectrum has more features than standard HBARs. We observe frequency beatings in the spectrum and there seem to be parasitic acoustic modes in between the main HBAR modes. We observed very similar acoustic spectra in two independent YSO samples with independent transducer transfers, indicating that these features come from the intrinsic acoustic properties of the system. The low symmetry of this crystal and its strong anisotropy necessarily lead to quasi-shear and quasi-bulk acoustic modes that can be substantially far from pure modes \cite{Auld}. Upon transmission from the LiNbO$_3$ transducer to the acoustically trirefringent YSO sample, shear waves decompose into the eigen modes of YSO. We believe that this conversion into modes having different phase velocities, combined with possible imperfections for instance on parallelism, is responsible for the richer Fabry-Perot spectrum that we observe in YSO. 

Nonetheless, like for Si or CaWO$_4$, we find a region in the center of the HBAR spectrum where modes are over-coupled, while modes on the wings are under-coupled. To obtain a large contrast, we find a mode close to critical coupling at 4.754 GHz, as shown in Fig. \ref{Fig:YSO}\textbf{c}. Using this acoustic mode, we perform paramagnetic spin resonance for various magnetic field orientations. We orient the magnetic field in the (b, D1, D2) coordinate system shown in Fig. \ref{Fig:YSO}\textbf{d}, D1 and D2 being the optical extinction axes, commonly used as a reference frame for the magneto-optical properties. The spin resonance data in the resulting three perpendicular planes are shown in Fig.\ref{Fig:YSO}\textbf{e, f, g}. Each substitution site for erbium in YSO has two magnetically non-equivalent orientations, related by a C2 symmetry parallel to the b axis \cite{GuillotNoel2006}. Consequently, in the (b, D1) and (b, D2) plane, we observe the expected four spin resonances, corresponding to the two crystallographic sites and the symmetry with respect to b \cite{Sun2008}. In the (D1, D2) plane, the nearly degenerate response of the two magnetically inequivalent orientations shows that the coordinate system of our magnetic field is well aligned with the crystal. \\

\begin{figure*}[ht]
\includegraphics{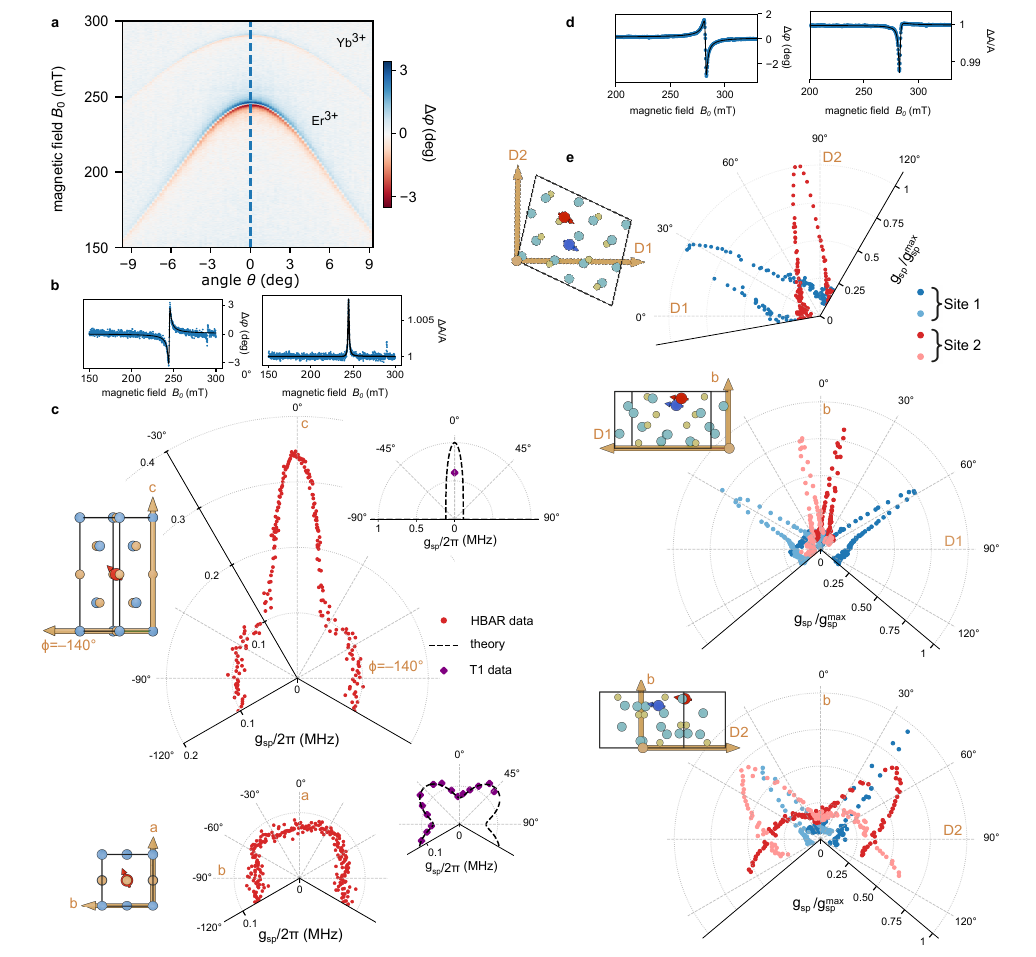}% Here is how to import EPS art
\caption{ 
\label{Fig:Couplage}
\textbf{Spin-phonon coupling for Er$^{3+}$ dopants in calcium tungstate (CaWO$_4$) and yttrium orthosilicate (Y$_2$SiO$_5$).} \textbf{a,} Acoustic paramagnetic resonance with the under-coupled HBAR mode at 4.275~GHz shown in Extended Data Fig. \ref{Fig:Undercoupled_CaWO$_4$}, at low microwave power ($\approx -103$~dBm at the HBAR input), as a function of the external magnetic field amplitude $B_0$ and its angle $\theta$ with respect to the c axis of CaWO$_4$ (a-c plane). \textbf{b,} Line cut at $\theta=0$, showing raw data (dots) and fit using spin susceptibility (solid lines, see Methods). \textbf{c,} Spin-phonon coupling strength for Er$^{3+}$:CaWO$_4$, in polar coordinates, at $\phi=-140 \degree$ (top pannel) and $\theta=90 \degree$ (a-b plane, bottom pannel). Red dots are HBAR data, dashed lines are the approximate crystal-field theory (see Methods) and purple diamonds are estimated spin-phonon coupling extracted with $T_1$ data from \cite{LeDantec2021,Wang2023}. \textbf{d,} Typical APR signal with an under-coupled HBAR mode in YSO at 4.406~GHz. Dots are raw data, line is a fit (see Methods). \textbf{e,} Spin-phonon coupling strength for Er$^{3+}$:YSO, in polar coordinates, normalized to the maximum measured coupling $g_\text{sp}^\text{max}/2\pi= 1.5\pm 0.4$ MHz. Both substitution sites are shown, in the three orthogonal planes of b-D1-D2.}
\end{figure*}

\noindent \textbf{Spin-phonon interactions}\\
We turn to the extraction of the strength and the anisotropy of the coupling between the spin ensembles and the HBAR modes. As shown in the Methods, the impact of a fully polarized spin ensemble on the microwave response of an HBAR can be modeled by a spin-phonon susceptibility, which depends on the spin-phonon coupling $g_\text{sp}$, the Larmor frequency (controlled by the external magnetic field) and the inhomogeneous linewidth of the ensemble $\Gamma_\text{inh}$. This model can be understood intuitively: the linewidth of the spin-phonon resonances (e.g. in Fig. \ref{Fig:YSO}\textbf{e, f, g}) is approximately given by $\Gamma_\text{inh}$ and the contrast is approximately given by $\frac{4g_\text{sp}^2}{\Gamma_\text{inh}}\frac{\kappa_\text{ext}}{\kappa_\text{int}^2-\kappa_\text{ext}^2}$. Using modes with $\kappa_\text{ext}=\kappa_\text{int}$ therefore leads to large contrasts but also large systematic uncertainties. Consequently, we target under-coupled HBAR modes ($\kappa_\text{ext}<\kappa_\text{int}$) when acquiring data to extract $g_\text{sp}$. This choice also mitigates the effect of possible Fano interferences \cite{Rieger2023}.  \\

We now perform APR measurements using the under-coupled mode shown in Extended Data Fig. \ref{Fig:Undercoupled_CaWO$_4$}. We also lower the microwave power to $\approx -103$~dBm at the HBAR input to stay away from any non-linearities (see supplementary information). The results are shown in Fig. \ref{Fig:Couplage}\textbf{a}. As expected for an under-coupled mode, we find a much lower phase contrast. We still clearly identify two spin resonances, associated to Er$^{3+}$ and Yb$^{3+}$ ions with zero nuclear spin. As shown in Fig. \ref{Fig:Couplage}\textbf{b}, we fit the phase response (acoustic dispersion) and amplitude response (acoustic dissipation) of the HBAR using the spin susceptibility model given in the Methods. In order to extract the spin-phonon coupling strength and anisotropy, we run this fit for every magnetic field orientation. Close to the c axis, we verify that our extraction of $g_\text{sp}$ is accurate by acquiring complete maps of the spin-phonon resonance as a function of probe frequency (see supplementary information). The full angular dependence of $g_\text{sp}$ for Er:CaWO$_4$ is shown in polar coordinates in Fig. \ref{Fig:Couplage}\textbf{c}. It reaches its maximum value when the magnetic field is aligned with the c axis of CaWO$_4$. For this orientation, we extract a spin-phonon cooperativity $C\approx0.52$ (see supplementary information). 

We observe a substantial anisotropy, that we qualitatively compare with a spin-lattice relaxation model and with existing experimental data on the relaxation time $T_1$ measured at dilution fridge temperature \cite{LeDantec2021,Wang2023}, shown in the insets of Fig. \ref{Fig:Couplage}\textbf{c}. As discussed in the Methods, our approximate model is based on crystal field theory. It is not sufficient to predict $g_\text{sp}$ quantitatively but it provides a trend in the anisotropy that agrees reasonably well with our data. Our measurements can serve as a basis for more advanced modeling, including the very recent effort based on inelastic neutron scattering analysis \cite{Sayat2026}. We also compare with existing $T_1$ data, using Fermi's golden rule and using the estimated 20 ppm concentration of erbium to estimate an average spin-phonon coupling strength (see Methods). Finally, it is worth recalling that the measured coupling rate $g_\text{sp}$, between the spin ensemble and the HBAR, is independent of the acoustic mode volume (as opposed to the coupling to a single spin). It only depends on the Er concentration. Nevertheless, we can estimate that there are around $10^{11}$ spins in the HBAR. \\ 

In YSO, as discussed previously, the acoustic spectrum is more complex than in the other materials that we have tested. The presence of multiple surrounding resonances leads to a background that also interacts with the spin ensemble. This phenomenon produces interferences on the spin resonances, even for seemingly very under-coupled modes (see supplementary information). An example of such spin resonance signal is shown in Fig. \ref{Fig:Couplage}\textbf{d}. We can fit this data by adding a Fano parameter to our acoustic model, but this additional free parameter adds significant uncertainty in the extraction of $g_\text{sp}$. To evaluate this uncertainty, we take a collection of HBAR modes on two different YSO samples. For each of these modes, we extract the full angular dependence $g_\text{sp}$. The results are shown in Extended Data Fig. \ref{Fig:extended_YSO}. We find that, within experimental noise, all sets of data show the same anisotropy, they match up to a global constant. In Fig. \ref{Fig:Couplage}\textbf{e} we show the full angular dependence of $g_\text{sp}$ for all Er sites in YSO, re-normalized by the maximum value $g_\text{sp}^\text{max}$. Based on the independent sets of data, we estimate $g_\text{sp}^\text{max}/2\pi= 1.5\pm 0.4$~MHz. This maximum coupling is obtained in the D1-D2 plane for $\phi= 81$~\degree. At this magnetic field orientation, we extract an HBAR-spin-ensemble cooperativity $C=0.43 \pm 0.21$ (see supplementary information).

The anisotropy of $g_\text{sp}$ is strong for Er:YSO, and it is notably not equivalent for the two substitution sites. This difference is particularly visible in the (b-D2) plane, where $g_\text{sp}$ for site 1 is substantially more anisotropic. A previous study \cite{Probst2013} has measured a $T_1\approx 4.3$ s at 20 mK in this system. Using Fermi's golden rule, this relaxation time translates into an average spin-phonon coupling rate $\approx 0.6$ MHz for 200 ppm of erbium. This order of magnitude matches the coupling rates $g_\text{sp}$ that we measure here. We are not aware of other experimental $T_1$ data on this system. The spin-lattice theoretical estimate that we proposed for CaWO$_4$ is much more difficult here, given the difficulty of modeling the crystal field in C1 symmetry. We have chosen not to propose this calculation for YSO, which would remain qualitative in nature, despite recent results on the estimation of the 27 parameters for site 1 \cite{Horvath2019}.  \\

\noindent \textbf{Discussion}\\
In general, the coupling $g_\text{sp}^0$ between a single spin and any acoustic resonator can be written as 

\begin{equation} \label{eq:dijEpsilon}
    g_\text{sp}^0=\sum_{ij}d_{ij} \epsilon_{ij}^\text{zpf}
\end{equation}

where $\epsilon_{ij}^\text{zpf}$ are the strain zero-point fluctuations of the resonator, with $i$ and $j$ running over the spatial coordinates $(x,y,z)$ of the strain tensor and $d_{ij}$ is the spin susceptibility to the corresponding strain direction. At the microscopic scale, the $d_{ij}$ coefficients come from the combined effects of spin-orbit coupling and the crystal field dependence on strain \cite{Larson1966,Whiteley2019,Meesala2018} (see also the Methods). Then the coupling to a spin ensemble scales as $g_\text{sp}=g_\text{sp}^0\sqrt{N}$, with $N$ the number of spin in the ensemble. 

From a fundamental perspective, one advantage of HBARs over other acoustic resonators such as surface acoustic wave (SAW) resonators, is the possibility to have pure acoustic polarization. In this case, coupling to the HBAR isolates a single term in the sum of equation \ref{eq:dijEpsilon}. By measuring samples with three different crystal cuts, and by using multimode piezoelectric transducers, it is possible to recover the six different $d_{ij}$ independently. Suitable multimode transducers could made by replacing X-cut LiNbO$_3$ with Y-cut LiNbO$_3$ (with shear and longitudinal modes at separate frequencies) and following the rest of our procedure. As discussed earlier, achieving pure acoustic polarization is challenging in complex materials such as YSO. However, it is realistic in many other materials with higher symmetries such as silicon or other scheelites (YVO$_4$, YLiF$_4$ for example). For our results in the sheelite CaWO$_4$, although we did not align our transducer with the target crystal, our HBAR modes mostly have strain components $\epsilon_{ab}$ and $\epsilon_{bc}$. \\ 

Finally, we note that there is a correlation between $g_\text{sp}$ and effective gyromagnetic ratio (i.e. g-tensor values). As shown in Fig. \ref{Fig:CaWO$_4$}\textbf{e} and \ref{Fig:YSO}\textbf{e,f,g}, large spin-phonon couplings are correlated with minimal values of gyromagnetic ratios. There are exceptions however, such as for the site 2 of YSO in the b-D2 plane (see red data points in Fig. \ref{Fig:Couplage}\textbf{e} and Fig. \ref{Fig:YSO}\textbf{g}).  These observations call for further theoretical works in modeling the wavefunctions. \\

\noindent \textbf{Conclusion and perspectives}\\
We have reported a new APR technique that can be implemented on mm-sized chips of any material as long as they can withstand double-side polishing and temperatures up to 150\degree C. Using this technique, we extracted the strength and the anisotropy of $g_\text{sp}$ for dilute erbium ion ensembles in CaWO$_4$ and YSO. YSO, with low symmetry and scarce data, illustrates the utility of our method. So far we have investigated Kramers ions, i.e. half-integer spins, and in particular erbium dopants for which we extracted $g_\text{sp}$. Erbium ions were good candidates to illustrate our method because they are easy to identify unambiguously in a vector magnetic field due to their specific g-factor, and were expected to have intermediate coupling strength. Non-Kramers ions (integer spins) are expected to have much larger spin-phonon coupling rates in high symmetry crystals, when the crystal field does not fully lift the multiplets degeneracy \cite{Larson1966}. Within the rare-earth dopants, holmium in CaWO$_4$ or praseodymium in ethyl sulfates are examples known to have nearly-degenerate ground state doublets \cite{Kirton1965,Larson1966} and are expected to have very large spin-phonon coupling. Future lines of work also include dopants in silicon \cite{Mansir2018} or diamond \cite{Meesala2018}, as well as magnetic materials \cite{Hiraishi2025}.

The erbium-doped materials we have measured show spin-phonon cooperativites approaching unity. This figure of merit shows that erbium-doped or ytterbium-doped HBARs could be used for microwave-to-optics transduction \cite{Han2026}, possibly using other materials such as YVO$_4$ \cite{Xie2025} or GdVO$_4$ \cite{Hiraishi2025}. In HBARs, both the optical wavelength and the acoustic wavelength at GHz frequencies are on the same order of magnitude (1 \textmu m). This property makes HBARs particularly well-suited devices to maximize the overlap between the spin ensemble and the optical and acoustic modes \cite{Yoon2023}. 

Finally, on the way towards coherent spin-phonon coupling with single spins, our APR technique is a strong asset to search for alternative to SiV centers in diamond \cite{Lee2017,Lemonde2018}. Once a good ion-matrix candidate is identified with APR on bulk samples, efforts can focus on developing the appropriate fabrication techniques to interface wavelength-scale acoustic nano-structures in this material \cite{Hugot2026,Chai2025}. \\

\noindent \textbf{Methods}\\ \nopagebreak
\noindent \textbf{Fabrication of the piezoelectric transducers}\\ \nopagebreak
The piezoelectric transducers are fabricated in batches, independently of the target crystal containing the spins. We start from commercial wafers of thin-film X-cut LiNbO$_3$ (500 nm) on high resistivity silicon (500 \textmu m). Using ion beam etching, we first thin down the LiNbO$_3$ film to around 250 nm and then pattern it to obtain the transducer disk, the tethers, etc. Then we perform partial laser etching of the silicon backside, with a depth of about 400~\textmu m deep. This step enables easier release of the LiNbO$_3$ structures, which are relatively large, at the end of the process. The next step is to pattern the top electrode and the CPW waveguide using lithography and a liftoff of Ti (5 nm) and Au (35 nm). To allow connection between the bottom electrode and the front-side circuits at a later step, we pattern wires going from the ground of the CPW to specific locations nearby the transducer disk. At these locations, the front-side metallic film intentionally goes over the edge of LiNbO$_3$ and onto silicon. We then release the LiNbO$_3$ structures using XeF2 gas etching, which etches from the backside as well as from the frontside, through the holes of the patterned LiNbO$_3$. At this point, the front-side metallic film parts that are over the edge of LiNbO$_3$ are also suspended, forming flaps. Finally we evaporate Ti (5 nm) and Au (35 nm) from the backside. This creates the bottom electrode of the transducer disk. This electrode connects to the CPW ground electrodes at the location of the flaps. These locations are visible in Fig. \ref{Fig1}\textbf{h}. \\

\noindent \textbf{Target crystal preparation}\\ \nopagebreak
The crystal axes of our samples are first aligned using X-ray diffraction measurements. We then dice the material into $4\times3$~mm samples, typically 300~\textmu m thick, and polish them on both sides. Parallelism of the polished faced is ensured via auto-collimation measurements during the polishing process. The CaWO$_4$ and YSO samples shown in this work have a surface roughness on the order of 1 to 2 nm.  In comparison, the surface roughness of our LiNbO$_3$ films is around 0.5 nm and that of our metallic electrodes on LiNbO$_3$ is around 1 nm. \\

\noindent \textbf{Acoustic properties of the LiNbO$_3$ transducer and the HBARs}\\ \nopagebreak
To understand how HBAR modes couple to propagating microwave signals, it is instructive to start by considering the properties of the free-standing piezoelectric transducer (i.e. in the absence of the crystal, as in Fig. \ref{Fig1}\textbf{b}). The transducer disk is well approximated by a parallel-plate capacitor (thickness around 250 nm, diameter 50 \textmu m), filled with a piezoelectric dielectric material and with a uniform electric field perpendicular to the disk plane. In its microwave response, i.e. in its electrical admittance, this capacitor has a piezoelectric contribution, with a $\lambda/2$ acoustic resonance close the frequency $f_0$ mentioned in the main text. Alone, this transducer is analogous to a thin-film bulk acoustic wave resonator (FBAR). Once transfered on the target crystal, the transducer couples to its overtone modes and mediates the coupling $\kappa_\text{ext}$ between these modes and the microwave line. \\

One can legitimately expect additional acoustic losses from the presence of the thin layer of PMMA that is used to glue the transducer. However, the participation ratio of this layer to the total HBAR is very small ($<2\times10^{-4}$). The PMMA layer can affect another relevant parameter of HBARs that is the frequency band over which acoustic modes have large $\kappa_\text{ext}$. In this case, it is the acoustic impedance mismatch of the PMMA that matters \cite{Gokhale2020}. However, as we aim to measure arbitrary crystals, we cannot target acoustic impedance matching between the transducer (LiNbO$_3$) and the crystal. In practice, we find that HBAR signals are typically faint at room temperature but increase considerably at low temperature, indicating that the PMMA acoustic transparency improves. On the materials we have tested for spin-phonon measurements, we have usefull HBAR modes over bandwidths on the order of $0.1 - 1$~GHz (see Fig. \ref{Fig:CaWO$_4$}\textbf{b}, Fig. \ref{Fig:YSO}\textbf{b} and Extended Data Fig. \ref{Fig:HBAR-Si} and \ref{Fig:Kext_CaWO$_4$}). \\

\noindent \textbf{Microwave response and spin-phonon susceptibility}\\ \nopagebreak
To obtain the spin-dependent microwave response of the HBAR modes, we write the standard coupled equations of motion for a harmonic mode coupled to spin ensemble. Using the input-output relations and the semi-classical decoupling \cite{Milburn}, we get the usual form for the microwave reflection coefficient of the HBAR :

\begin{equation} \label{eq:S11}
    S_{11}  = 1 - \frac{\kappa_\text{ext}}{\frac{\kappa_\text{tot}}{2} - i(\omega_m - \omega) + i \chi_{sp}} 
\end{equation}

where $\chi_{sp}=\frac{g_\text{sp}^2}{\omega_s-\omega+i \Gamma_\text{inh}/2}$ is the spin-phonon susceptibility of the polarized spin ensemble. As discussed in the main text, in the limit of weak coupling ($g_\text{sp}^2<\Gamma_\text{inh} \kappa_\text{tot}$), we get a first order approximation for the phase and amplitude contrasts $\Delta\varphi$ and $\Delta A/A$ on a spin-phonon resonance:

\begin{equation}
    \Delta \varphi \approx 4 \Re(\chi_{sp}) \frac{\kappa_\text{ext}}{\kappa_\text{int}^2-\kappa_\text{ext}^2} 
\end{equation}
\begin{equation}
    \frac{\Delta A}{A} \approx 4 \Im(\chi_{sp}) \frac{\kappa_\text{ext}}{\kappa_\text{int}^2-\kappa_\text{ext}^2} 
\end{equation}

This result shows the interest of choosing a critically coupled mode ($\kappa_\text{ext} \approx \kappa_\text{int}$) to maximize the microwave signal, while choosing an under-coupled mode ($\kappa_\text{ext} < \kappa_\text{int}$) to extract $\chi_{sp}$ with minimal systematic uncertainty. As discussed in the main text and in the supplementary information, some HBAR modes have a substantial asymmetry in their frequency response. We take this into account by adding replacing $\kappa_\text{ext}$ with $\kappa_\text{ext}e^{i\phi_F}$ in equation \ref{eq:S11}, and we use this equation to extract $g_\text{sp}$. \\

\noindent \textbf{Approximate coupling $g_\text{sp}$ from spin-lattice relaxation theory in CaWO$_4$}\\ \nopagebreak
The question of phonon coupling arose very early on in order to explain the limitations of the spin lifetimes by spin-lattice relaxation. Jeffries at al. describe how to extract an average coupling coefficient from knowledge of the crystal field parameters \cite{Larson1966}. The method is explained in detail in ref. \cite{Scott1962}. The square root of equation (16) therein, that we calculate for CaWO$_4$, directly provides an estimate of $g_\text{sp}$ for an average strain field (average value of the $d_{ij}$ in equation \ref{eq:dijEpsilon}). This can be compared with experimental data. Since the crystal field parameters are known \cite{Enrique1971}, we also use the series of normalization factors calculated in Table I of ref. \cite{Scott1962}. \\

\noindent \textbf{Order of magnitude for $g_\text{sp}$ from $T_1$ data for Er$^{3+}$:CaWO$_4$}\\ \nopagebreak
Here we follow the general reasoning of ref. \cite{Abragam1970}. We start by making the very rough approximation that spin-lattice relaxation can be modeled by the decay into a single acoustic mode, with an interaction Hamiltonian of the Jaynes-Cummings type. For a single spin we write a simplified version of equation \ref{eq:dijEpsilon} for the Jaynes-Cummings coupling constant:

\begin{equation} \label{eq:Simplified-JC-coupling}
    g_\text{sp}^0 = d_s \epsilon_\text{zpf}
\end{equation}

where $d_s$ is an average spin-strain sensitivity (purely a characteristic of the spin) and $\epsilon_\text{zpf}$ is the strain zero-point fluctuations of the acoustic mode considered. For Kramers ions like erbium, $d_s$ depends on the spin mixing induced by Zeeman splitting on the ground state Kramers doublet. Following ref. \cite{Abragam1970}, we use the perturbative expression linear in frequency, $d_s=d_{s0}\omega_s$. We also write $\epsilon_\text{zpf} \approx \sqrt{\frac{\hbar\omega_a}{2\rho V v_\varphi^{2}}}$, with $\omega_a/2\pi$ the acoustic mode frequency, $\rho$ the density of the material, $V$ the acoustic mode volume and $v_\varphi$ is the acoustic phase velocity. We then use Fermi's golden rule, when the acoustic mode is resonant with the spin Larmor frequency $\omega_s$:

\begin{equation} \label{eq:FermiGoldenRule}
    \frac{1}{T_1}= \frac{\hbar}{2\pi} \frac{d_{s0}^2 \omega_s^5}{\rho v_\varphi^5 }
\end{equation}
This result is independent of the volume of acoustic mode and recovers the known $\omega_s^5$ dependence. We use this result to extract $d_{s0}$ from $T_1$ data measured at mK temperatures and GHz Larmor frequencies in \cite{LeDantec2021,Wang2023}. Finally, with the spin concentration $C_0$ in the HBAR, we use again equation \ref{eq:Simplified-JC-coupling} to get an estimate of the coupling to the spin ensemble:
\begin{equation} 
    g_\text{sp}= d_{s0} \omega_s \sqrt{\frac{\hbar\omega_a C_0}{2\rho v_\varphi}}
\end{equation}
where the frequencies here are that of our experiment. \\

\noindent \textbf{Data availability} \\ 
The data supporting the findings of this paper is available from the corresponding authors upon request. \\

\noindent \textbf{Acknowledgments} \\
We acknowledge fruitful discussions with P. Bertet, A. Bienfait, A. Reinhardt and T. Baron. We are indebted with Eric Eyraud and Wolfgang Wernsdorfer for their support with the dilution fridge used for this experiment. We thank R. Bouland and J. Debray for their help with the polishing of the samples and S. Le Denmat for surface roughness measurements. We thank all the members of the superconducting quantum circuits group at Neel for helpful discussions. The piezoelectric transducers were fabricated in the clean room facility of the Néel Institute and we thank all the clean room staff for their assistance.  This work was supported by the French National Research Agency (ANR) under the France 2030 plan, with reference ANR-22-PETQ-0003 and ANR-22-PETQ-0010 (QMEMO), as well as through the project MagMech (ANR-20-CE47-0004-01). S.Z. acknowledges financial support from QuanTEdu-France n°ANR-22-CMAS-0001 France 2030. \\

\noindent \textbf{Author contributions}\\
Q.G. and J.J.V. designed the experiment. Q.G. fabricated the devices with inputs and support from S.Z., J.J. and J.J.V.. Q.G. realized the measurements and performed the data analysis with the help of A.H., E. B. and J.J.V.. J.J. and L. D.-R. fabricated the superconducting vector magnetic field and parts of the cryogenic setup. T. C. calculated the approximate coupling from spin-lattice relaxation theory. J.J.V. supervised the project with help from F.B. and T.C.. Q.G., T. C. and J.J.V. wrote the manuscript with inputs from all authors. \\

\noindent \textbf{Competing interests}\\
The authors declare no competing interests. \\

\bibliography{BiblioGreffe2025}% Produces the bibliography via BibTeX.

\renewcommand{\figurename}{\textbf{Extended Data Fig.}}
\setcounter{figure}{0}

\begin{figure*}[ht]
\includegraphics{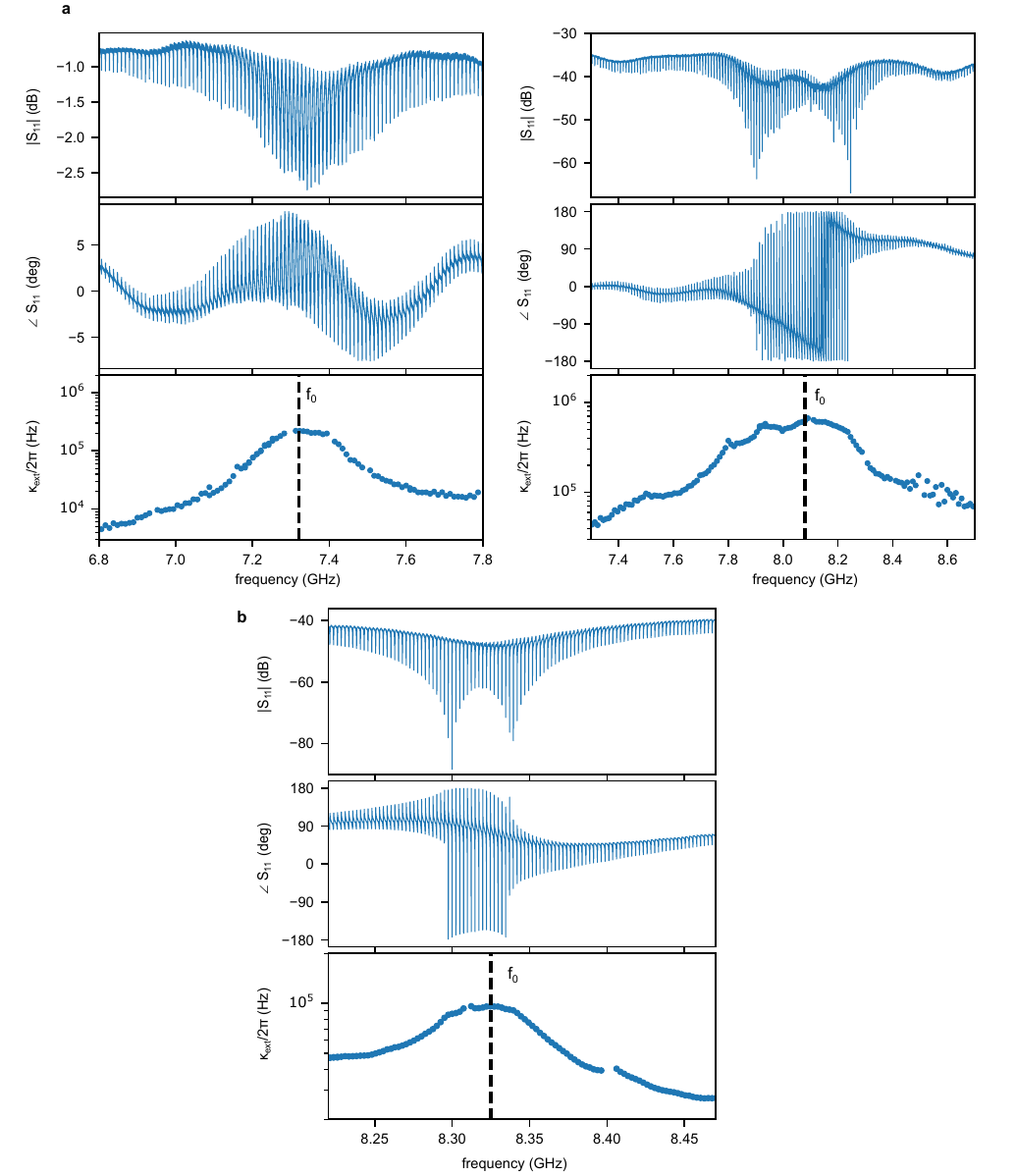}% Here is how to import EPS art
\caption{
\label{Fig:HBAR-Si}
\textbf{Manufacturing HBAR with large external coupling $\kappa_\text{ext}$}
\textbf{a,} Microwave reflection $S_{11}$ (magnitude and phase) and extracted external coupling $\kappa_\text{ext}$ for two HBARs realized on Si [100] and measured at low temperature. The frequencies of operation are larger here than for the HBAR presented in the main text because the transducers used on these samples where made with aluminum electrodes, instead of gold. \textbf{b,} Microwave reflection $S_{11}$ and extracted external coupling $\kappa_\text{ext}$ for an HBAR realized on glass (Borofloat).
}
\end{figure*}

\begin{table*}[t]
\renewcommand{\arraystretch}{1.4}
\centering
\caption{
\label{Table:MeasuredHBARs}
\textbf{Summary of measured HBAR devices that were fabricated using the visco-elastic transfer technique.} The table provides the approximate maximal external coupling values $\kappa_\text{ext}^\text{max}$ extracted for HBAR modes close to $f_0$.}
\begin{tabular}{|c|c|c|}

\hline
\parbox{2.5cm}{\centering \textbf{ \newline Target \newline material \newline  }} &
\parbox{5cm}{\centering \textbf{ \newline Transducer diameter (µm), \newline electrode material \newline }} &
\parbox{3.5cm}{\centering \textbf{ \newline $\kappa_\text{ext}^\text{max}$/2$\pi$ (MHz) \\ close to $f_0$ \newline }} \\ \hline

\multirow{5}{*}{ \centering Si} & 100 µm, Al & 0.06 \\ \cline{2-3}
& \multirow{4}{*}{50 µm, Al} & 0.2 \\ \cline{3-3}
& & 0.3 \\ \cline{3-3}
& & 0.4 \\ \cline{3-3}
& & 0.7 \\ \cline{2-3}
& 50 µm, Au & 0.3 \\ \hline
Borofloat & 15 µm, Al & 0.09 \\ \hline
\multirow{5}{*}{CaWO$_4$} & 100 µm, Al & 0.1 \\ \cline{2-3}
& \multirow{3}{*}{50 µm, Al} & 0.3 \\ \cline{3-3}
& & 0.5 \\ \cline{3-3}
& & 0.4 \\ \cline{2-3}
& 50 µm, Au & 0.5 \\ \hline
\multirow{2}{*}{Y$_2$SiO$_5$} & \multirow{2}{*}{50 µm, Au} & 0.3 \\ \cline{3-3}
& & 0.2 \\ \hline
\end{tabular}
\end{table*}

\begin{figure*}[ht]
\includegraphics{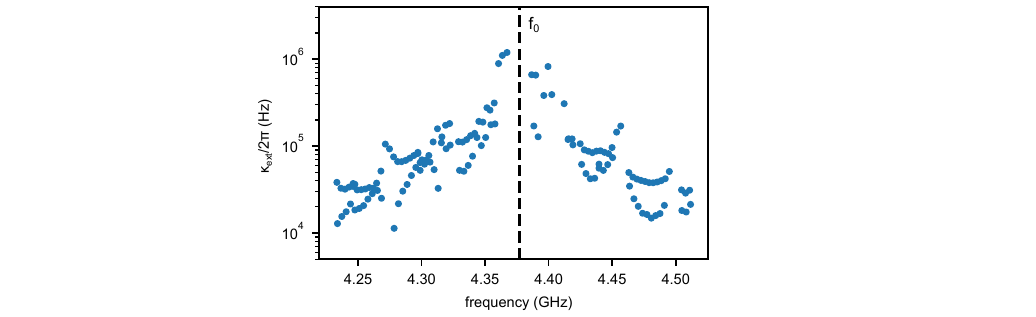}% Here is how to import EPS art
\caption{
\label{Fig:Kext_CaWO$_4$}
\textbf{External coupling $\kappa_\text{ext}$ for the CaWO$_4$ sample shown in Fig. \ref{Fig:CaWO$_4$}}}.
\end{figure*}

\begin{figure*}[ht]
\includegraphics{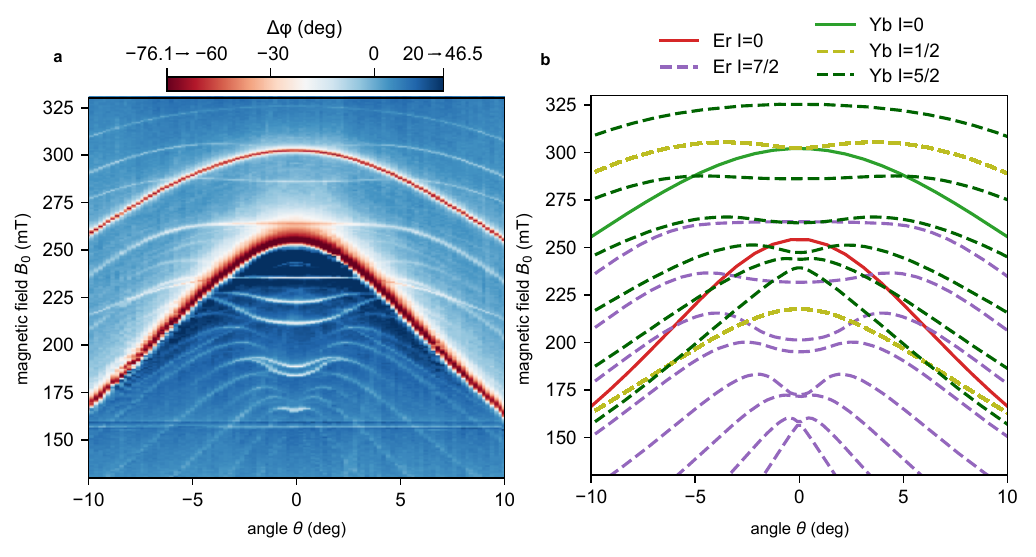}% Here is how to import EPS art
\caption{
\label{Fig:ZoomCaWO$_4$-c-axis}
\textbf{APR of CaWO$_4$, zoom around the c-axis}
\textbf{a,} Measured relative phase shift of the microwave reflection on the HBAR (color scale), as a function of the external magnetic field amplitude $B_0$ and angle $\theta$ with respect to the c axis. This data is obtained with a critically-coupled HBAR mode and large microwave power ($-72$~dBm at the input of the HBAR). \textbf{b,} Calculated spin resonance lines based on tabulated values for the g-tensor and hyperfine coupling of Er$^{3+}$ and Yb$^{3+}$ isotopes in CaWO$_4$ \cite{Bertaina2007,Sattler1970}.}
\end{figure*}

\begin{figure*}[ht]
\includegraphics{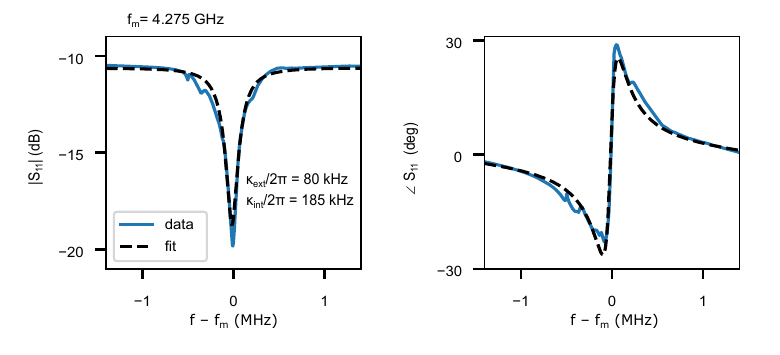}% Here is how to import EPS art
\caption{
\label{Fig:Undercoupled_CaWO$_4$}
\textbf{HBAR mode in CaWO$_4$ used to extract $g_\text{sp}$ in Fig. \ref{Fig:Couplage}.} 
}
\end{figure*}

\begin{figure*}[ht]
\includegraphics{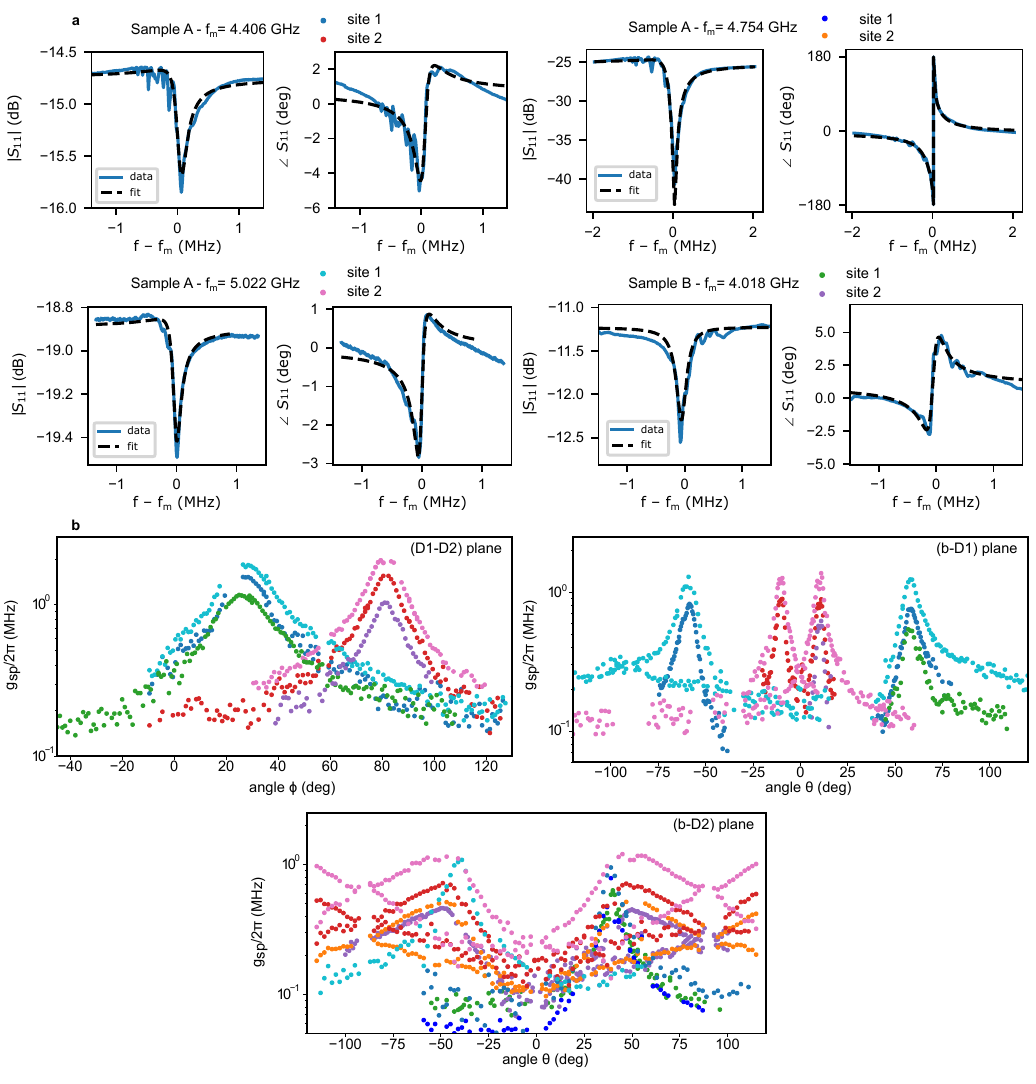}% Here is how to import EPS art
\caption{
\label{Fig:extended_YSO}
\textbf{HBAR modes and $g_\text{sp}$ data used to extract the YSO data shown in Fig. \ref{Fig:Couplage}.}
\textbf{a,} Microwave reflection measurements on four different modes, from two different samples. Solid lines are raw data and dashed lines are fits (see Methods). \textbf{b,} Spin-phonon coupling $g_\text{sp}$ for both YSO sites, extracted for each mode shown in \textbf{a}, in the three orthogonal planes of b-D1-D2. This shows that all data sets have the same anisotropy. From this data we estimate the uncertainty of the absolute value of the coupling and extract $g_\text{sp}^\text{max}/2\pi= 1.5\pm 0.4$~MHz. 
}
\end{figure*}

\clearpage

\begin{widetext}
 \centering
 \large \bfseries{Supplementary information for: A material-agnostic platform to probe spin-phonon interactions using high-overtone bulk acoustic wave resonators}\\[1.5em]
\end{widetext}

\title{Supplementary Information for: A material-agnostic platform to probe spin-phonon interactions using high-overtone bulk acoustic wave resonators}% Force line breaks with \\

\maketitle
%\onecolumngrid  % ← Passe en une seule colonne
\tableofcontents
\setcounter{figure}{0}
\renewcommand{\thefigure}{S\arabic{figure}}
%\section*{SUPPLEMENTARY MATERIAL}  % Titre en majuscules et non numéroté

\section{Measurement setup and vector magnetic field calibration}

The samples are cooled down to $\sim$ 60 mK in a homemade dilution refrigerator. Microwave measurements are performed using a vector network analyzer (Keysight P9373A). The microwave input signal is attenuated by 68 dB from the VNA to the sample , while the output is amplified by 59 dB over the 4-12 GHz band. \\

\begin{figure}[ht]
\includegraphics{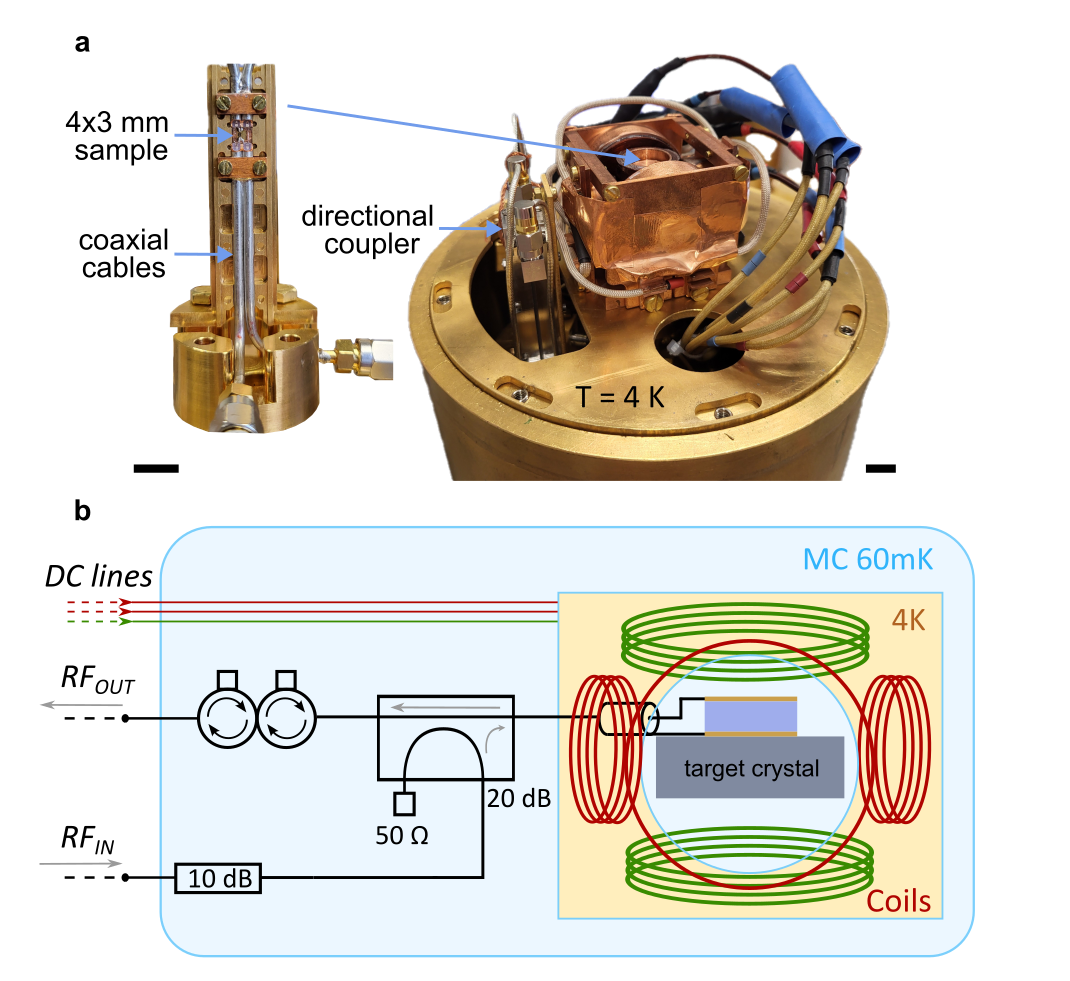}% Here is how to import EPS art
\caption{\textbf{a,} Photograph of the sample holder thermally anchored on the mixing chamber stage (MC) of the dilution fridge (left) and photograph the vector magnet (right). Scale bar is 10 mm. \textbf{b,} Schematic of the cryogenic microwave measurement setup.}
\label{microwave_setup}
\end{figure}

To generate the magnetic field, we use a homemade three-dimensional vector magnet composed of a solenoid coil surrounded by two orthogonal split coils. The coils are designed to deliver a magnetic field of up to 1 T and to accommodate a sample holder for a $4 \times 3$~mm sample. The coils are mounted on the 4 K stage of the cryostat. Fig \ref{microwave_setup} (a) shows a photograph of the 3D vector magnet and sample holder and Fig \ref{microwave_setup} (b) presents a schematic of the microwave setup installed on the mixing chamber. To calibrate the three-dimensional vector magnetic field, we first performed room temperature measurements using a Hall probe. To calibrate the field further, we relied on the electron spin resonance of Er $(I=0)$ transition in CaWO$_4$. Given the known crystal axes and the gyromagnetic tensor, this allows us to quantify the magnetic field strength generated by each coil.

\section{Linear regime}
\subsection{Probe power to extract $g_\text{sp}$}

When acquiring data for the purpose of extracting the spin-phonon coupling $g_\text{sp}$, we adjust the probe power in order to remain in a regime where the microwave response is independent of power. \\

\begin{figure}[h!]
    \centering
    \includegraphics{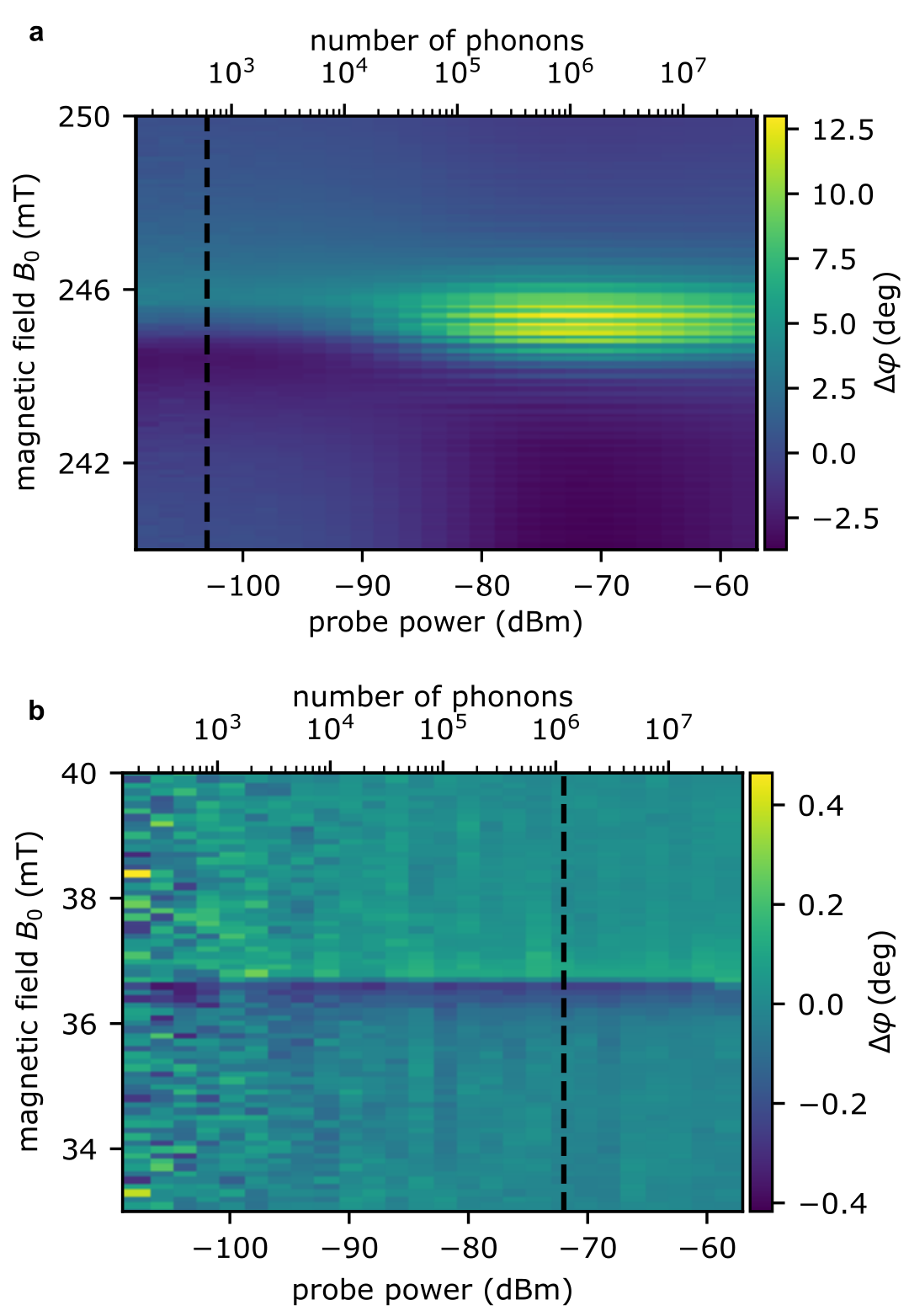}% Here is how to import EPS art
    \caption{Power dependence of the Er $(I=0)$ line in CaWO$_4$, measured with an HBAR mode at 4.275 GHz. Both panels show the microwave phase shift $\Delta\varphi$ as a function of the probe power at the HBAR input, for varying magnetic fields. The field is along the crystal c axis in \textbf{a} and in the a-b plane at $\phi =-140 \degree$ in \textbf{b}. The dotted black line corresponds to the power used in the main text: $-103$ dBm and $-72$ dBm near the c-axis and in the a-b plane respectively.
    }
    \label{fig:S1:resonance_power}
\end{figure}

Fig \ref{fig:S1:resonance_power} shows the power dependence of the phase of $S_{11}$ in CaWO$_4$, measured around the spin-phonon resonance along the c axis.  For moderate incident powers, the measured phase shift is independent of power, and is consistent with Eq \ref{eq:S11_fano}. At large powers, the response becomes non-linear and prevents us from extracting $g_\text{sp}$. To maximize the the signal-to-noise ratio while staying in the linear regime, we adapt the probe power to the magnetic field orientation, we use $-103$ dBm around the c-axis ($-10\degree<\theta<10\degree$) and $-72$ dBm elsewhere. \\

In YSO, the microwave response was measured as a function of power in the (D1, D2) plane where the coupling is maximal for both substitution sites. The response of the spin-phonon resonances of Er $(I=0)$ spins was independent of power up to $-72$ dBm, which is the power used to acquire the data shown in the main text.

\subsection{HBAR quality factor independent of power} \label{subsec:indep-Q}

\begin{figure}[ht]
\includegraphics{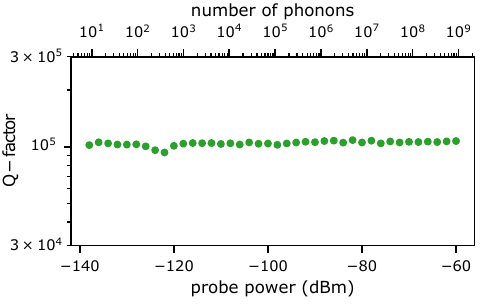}% Here is how to import EPS art
\caption{Power dependence of high quality factor acoustic mode at 5.075 GHz in the CaWO$_4$ sample discussed in the main text.
}
\label{Qfactor}
\end{figure}

We also characterized, as a function of power, the quality factor $Q$ of HBAR modes in CaWO$_4$ and YSO crystals. Fig \ref{Qfactor} shows such a measurement for one of the highest $Q$-factor modes in the CaWO$_4$ crystal (5.075 GHz, under-coupled mode). Measurements are performed at $B_0 = 10$ mT to break the superconductivity of the bonding wires. $Q$ is independent of power on a wide range from few phonons to $10^9$ phonons in the acoustic mode. This behavior was also observed in an over-coupled (lower $Q$) mode. The HBAR modes in YSO were also independent of the input power. \\

\section{Extraction of $g_{\text{sp}}$ and $\Gamma_{\text{inh}}$}

\begin{figure*}[ht]
    \centering
    \includegraphics{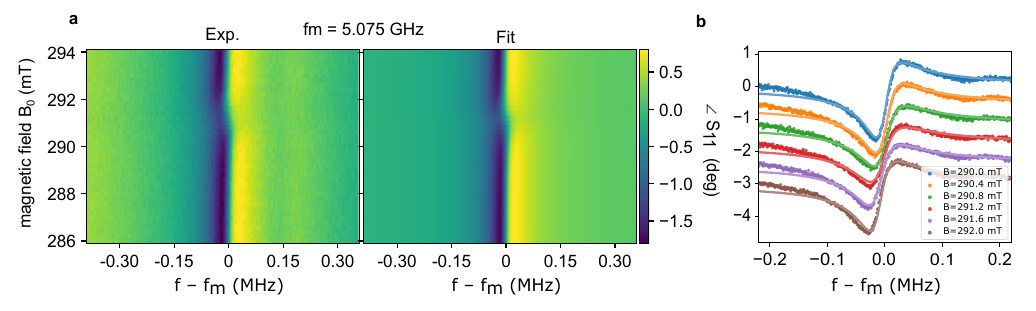}% Here is how to import EPS art
    \caption{Spin-phonon resonance of Er $(I=0)$ along the c-axis of CaWO$_4$. \textbf{a,} S$_{11}$ phase as function of probe frequency for a varying magnetic field. Left panel is experimental data and right panel is the result of the fit. The parameters obtained from the fit are shown in Table \ref{Table:parameter}. \textbf{b,} Line-cuts at magnetic fields near the spin-phonon resonance, dots are data and solid lines are fit results. 
    }
    \label{Fig:freq-B_CaWO4}
\end{figure*}

As discussed in the main text and in the Methods, to extract $g_{sp}$ and $\Gamma_{\text{inh}}$ from experimental data, we fit the microwave response of an HBAR mode at resonance with the spin ensemble using the model (see Methods):  

\begin{equation} \label{eq:S11_fano}
    S_{11}  = 1 - \frac{\kappa_\text{ext} e^{i\, \phi_F}}{\frac{\kappa_\text{int} + \kappa_\text{ext}e^{i\, \phi_F}}{2} - i(\omega_m - \omega) + i \chi_{sp}} 
\end{equation}

where $\chi_{sp}=\frac{g_\text{sp}^2}{\omega_s-\omega+i \Gamma_\text{inh}/2}$ is the spin-phonon susceptibility of the polarized spin ensemble, $\kappa_{\text{ext}}$ is the external coupling, $\kappa_{\text{int}}$ is the internal loss rate, $\phi_F$ captures the asymmetry of the HBAR response and $\omega_s=\mu_B g_\text{gyro} B_0$ is the Larmor frequency ($\mu_B$ is the Bohr magneton, $g_\text{gyro}$ the effective gyromagnetic ratio).

\subsection{CaWO$_4$}
For the CaWO$_4$ data shown in Fig. 4\textbf{c} of the main text, we extract $g_\text{sp}$ using an under-coupled HBAR mode with minimal asymmetry ($\phi_F\approx-0.09$, so $\kappa_\text{ext} e^{i \phi_F }\approx \kappa_\text{ext}$). Besides, we have checked that on modes with other $\phi_F$ values, we obtain very similar $g_\text{sp}$ values. To further support the validity of our fitting procedure, we acquire, for field orientations near the c axis, maps of the acoustic resonance as function of field strength. One of these maps is shown in Fig. \ref{Fig:freq-B_CaWO4}, for a high-$Q$ mode around 5.075~GHz. This measurement shows the expected behavior when the Larmor frequency $\omega_s$ is in resonance with the acoustic mode, at $B_0 = 291$~mT. We evaluate the bare HBAR parameters by fitting the off-resonance data at 286~mT and we evaluate $g_{\text{sp}}$ and the inhomogeneous linewidth $\Gamma_{\text{inh}}$ by fitting the entire map. The fit results are shown in the right panel of Fig. \ref{Fig:freq-B_CaWO4}\textbf{a} and in Fig. \ref{Fig:freq-B_CaWO4}\textbf{b}. The parameters obtained from the fit are shown in the Table \ref{Table:parameter}. The extracted coupling strength  $g_\text{sp}/2\pi =$ 0.36 MHz is consistent with the value reported in the main text for the c-axis (0.34 MHz). From this result, we also estimate the cooperativity between the HBAR mode and the spin ensemble as 

\[
C = \frac{ 4g_{sp}^2 }{ \Gamma_\text{inh} \left( \kappa_\text{int} + \kappa_\text{ext} \right) }
\]

which gives $C = 0.52$, as quoted in the main text.

\begin{table}[h!] 
\centering
\begin{tabular}{|c|c|}
\hline

\textbf{Parameter} & \textbf{Fit result} \\
\hline

$\omega_m / 2\pi$ & $5.0755$ GHz \\
$\kappa_\text{{int}}/ 2\pi$& $37  $ kHz \\
$\kappa_\text{{ext}}/ 2\pi$& $1  $ kHz \\
$\phi_F$ & $ 0.44 $ \\
$g_\text{{sp}}/ 2\pi$& $ 0.36  ~ $MHz \\
$\Gamma_\text{inh}/ 2\pi$& $ 25.5 $ MHz\\
$g_{\text{gyro}}$ & $ 1.245 $ \\

\hline
\end{tabular}
\caption{\label{Table:parameter} Parameters obtained by fitting the probe frequency - magnetic field map of Fig. \ref{Fig:freq-B_CaWO4}}
\end{table}

\subsection{YSO}

\begin{figure}[ht]
\includegraphics{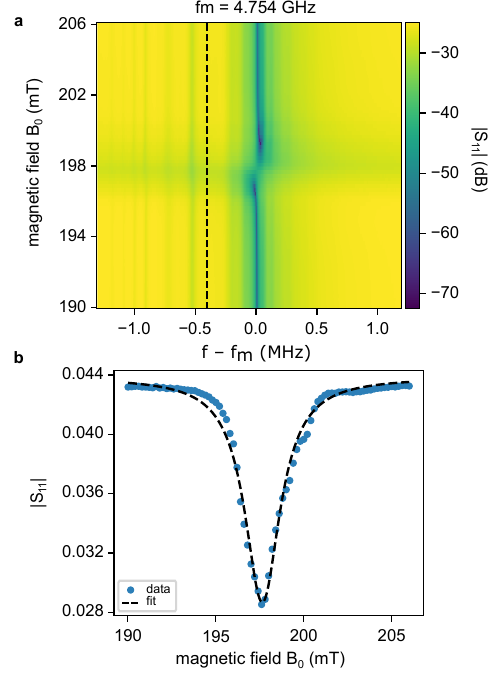}% Here is how to import EPS art
\caption{Spin-phonon resonance of Er $(I=0)$ in YSO, in the D1-D2 plane at $\phi=81 \degree$. \textbf{a,} $|S_{11}|$ as function of probe frequency and magnetic field. \textbf{b,} Spin ensemble lineshape measured at the probe frequency indicated by the black dashed line in \textbf{a}. Dots are data and dashed line is a lorenztian fit used to extract $\Gamma_\text{inh}$ at this magnetic field orientation.}
\label{Fig:freq-B_YSO}
\end{figure}

In YSO, as discussed in the main text, the more complex acoustic spectra make it difficult to extract $g_{\text{sp}}$ and $\Gamma_{\text{inh}}$ confidently. As a result, we decided to run our measurements on several HBAR modes, on two different YSO samples (see Extended Data Fig. 5). The results show  that the data sets taken on various HBAR modes are not identical, but they show the same anisotropy and they match up to a global constant. From this data we estimate the maximum coupling stength $g_\text{sp}^\text{max}/2 \pi = 1.5\pm 0.4 $ MHz. To extract the cooperativity with more confidence, we acquire the probe frequency dependence of the spin-phonon resonance where $g_{\text{sp}}$ is maximal (D1-D2 plane, site 2 at $\phi =81 \degree$). The result is shown in Fig. \ref{Fig:freq-B_YSO}\textbf{a}. We observe several additional acoustic modes and a background that interacts with the spin transition at 198 mT. In fact, the microwave contrast on the spin transition extends far from the main resonance at $f_m\approx4.754$ GHz. These contributions interfere altogether and the model of equation \ref{eq:S11_fano} cannot reasonably fit this data set, unlike that of CaWO$_4$ in Fig. \ref{Fig:freq-B_CaWO4}. However, from this measurement, we extract the inhomogeneous linewidth of this spin ensemble by fitting the data to a Lorentzian line-shape at a probe frequency detuned from the main HBAR resonance. This fit is shown in Fig. \ref{Fig:freq-B_YSO}\textbf{b} and gives $g_\text{gyro} = 1.72$ and a full width at half-maximum of 2.5 mT, corresponding to $\Gamma_\text{inh}/2\pi =  60$ MHz. By repeating this procedure for several frequencies, we obtain $\Gamma_\text{inh}/2 \pi = 62 \pm 4 $ MHz. Finally, the lowest HBAR linewidth we measure in YSO is $\kappa_\text{tot}/2\pi \approx 0.25$ MHz, but with the HBAR mode shown in Fig. \ref{Fig:freq-B_YSO}, we obtain $\kappa_\text{tot}/2\pi \approx 0.33$ MHz and we compute the cooperativity at $\phi = 81 \degree$ in the D1-D2 plane: $C = 0.43 \pm 0.21$.

\end{document}